\documentclass[prl,twocolumn,floatfix,superscriptaddress]{revtex4}
\usepackage[dvips]{graphicx}
\usepackage{epsfig}

\usepackage{verbatim}

\usepackage{subfigure}

\usepackage{amsmath}

\usepackage{natbib}

\begin{document}

\title{Microscopic Calculation of Flow Stress in Cu-Mg Metallic Glass}

\date{\today}

%81.05.Kf       Glasses (including metallic glasses)


\author{Nicholas P. Bailey}
\email{nbailey@fysik.dtu.dk}
\affiliation{CAMP, Department of Physics, Technical University of Denmark, 2800 Lyngby, Denmark and Materials Research Department, Ris{\o} National 
Laboratory, DK-4000, Roskilde, Denmark}

\author{Jakob Schi{\o}tz}
\author{ Karsten W. Jacobsen}
\affiliation{CAMP, Department of Physics, Technical University of Denmark, 2800 Lyngby, Denmark}

\begin{abstract}
We have carried out shear-deformation simulations on amorphous Mg-Cu
systems
at zero temperature and pressure, containing 2048-131072 atoms. At the
largest size a smooth stress-strain curve is obtained with a well-defined
flow stress. In the smallest system there are severe discontinuities in the
stress-strain curve caused by localized plastic events. We show that the
events can be characterized by a slip volume and a critical stress and we
determine the distribution of these quantities from the ensemble of all
events occurring in the small system. The distribution of critical stresses
at
which the enthalpy barriers for the individual events vanish is spread
between 200 MPa and 500 MPa with a mean of 316 MPa, close to the flow stress
observed in the largest system.
\end{abstract}

\maketitle

The mechanical properties of bulk metallic glasses\cite{Greer:1995,
Johnson:1999} (BMGs) are the subject of intense research. A host of
applications is envisaged if only reasonable macroscopic plasticity
could be achieved, rather than the intense localization into shear
bands which typically occurs. Detailed knowledge of plastic deformation 
mechanisms in glasses, and
their connection to macroscopic flow properties, however, remains
elusive: While in crystals the dislocation provides a well defined
starting point for estimates of flow stress, in glasses there is no
such easily characterizable defect.

Various recent theories\cite{Khonik/others:1998,
Langer/Pechenik/Falk:2003} take as a starting point the existence of a
collection of ``relaxation centers''\cite{Khonik/others:1998} or
``shear-transformation zones''(STZs)\cite{Falk/Langer:1998,
Langer/Pechenik/Falk:2003} which operate as localized centers of
deformation. Based on a set of assumptions about the operation of the
relaxation centers the total rate of plastic deformation is obtained
from the collective behavior of the local centers.

Indeed, several simulations of deformation in amorphous
metals\cite{Maeda/Takeuchi:1981, Srolovitz/Vitek/Egami:1982,
Falk/Langer:1998, Malandro/Lacks:1999, Lund/Schuh:2003b} have
established that the plastic behavior involves localized events with
up to 150 atoms\cite{Srolovitz/Vitek/Egami:1982}. The transforming
regions have been extensively studied from a potential energy
landscape point of view by Lacks\cite{Malandro/Lacks:1998,
Malandro/Lacks:1999, Lacks:2001}, and the response to shear and normal
stresses have been investigated for Lennard-Jones
systems\cite{Lund/Schuh:2003a}. In the latter work the connection
between the properties of a single model STZ and the yield stress was
discussed through the application of the Mohr-Coulomb yield
criterion. Systematic studies of the dependence of
macroscopic properties on the full ensemble of local deformation
events, however, are still needed.

\begin{figure}
\epsfig{file=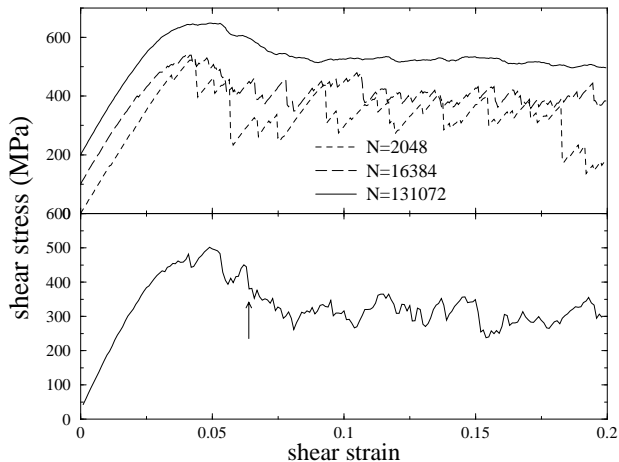,width=3.2 in}
\caption{\label{stressStrainCurves}Upper panel, stress-strain curves for 
different-sized systems undergoing pure shear at zero temperature, shifted for
 clarity. For the largest system the peak and flow stresses are 450MPa and 
320MPa respectively, or about $\mu/16$ and $\mu/23$ where $\mu$=7.3GPa is the 
shear modulus. Lower panel, stress-strain curve 
of a sub-volume of the 131072-atom system, selected because visualization (via 
deviation from affine deformation) indicated an event there at strain 0.064.}
\end{figure}

In this Letter, we describe zero temperature simulations of plastic
deformation of a Cu-Mg glass and demonstrate a direct
connection between the statistical properties of localized deformation events
and the flow stress at mesoscopic length scales (10--100 nm). Our main results
 are (i) while for small systems the stress-strain curve shows sharp
drops signifying individual deformation events, when the system size is over 
$10^5$ atoms, the stress strain curve becomes quite smooth, with a clear
flow stress of 320--330MPa (Fig.\ref{stressStrainCurves}); (ii) analysis of the
 individual events in the smallest system can be quantified in terms of two
quantities, a critical stress $\sigma_c$ and a quantity we call the
``slip volume'' $V_{\textrm{slip}}$; (iii) the mesoscopic flow stress is the
 mean of the distribution of $\sigma_c$, and the spatial density of 
transforming regions per unit strain in a large system is the inverse of the
 mean of $V_{\textrm{slip}}$.

The simulated material 
is Mg$_{0.85}$Cu$_{0.15}$, which is the optimal glass-forming composition for 
the Mg-Cu system\cite{Sommer/Bucher/Predal:1980}. This system is interesting 
because the addition of a small amount of Y makes it a BMG. The interatomic 
potential is the effective medium theory 
(EMT)\cite{Jacobsen/Stoltze/Norskov:1996}, fitted to properties of the pure 
elements and intermetallic compounds obtained from experiment and density 
functional theory calculations. Details of the potential and of the method for
 creating the zero-temperature glassy configurations may be found in
 Ref.~\onlinecite{Bailey/Schiotz/Jacobsen:2004}; for the 16384- and  
131072-atom systems we have used the cooling rate of 0.72 
K/ps. The main simulations involve continuous deformation of systems containing
 2048, 16384 and 131072 atoms. The nominally zero-temperature configurations 
resulting from the cooling
 runs were further minimized before deformation runs, with respect to both
atomic positions and the vectors describing the periodic supercell. The box
vectors are in fact controlled by six strain degrees of freedom, which also
play a role in the deformation simulations. We restrict to pure shear.
The deformation simulations are strain-controlled: the relevant component of 
strain is incremented in steps of 0.0005. After each step the remaining degrees
 of freedom---the atomic positions and other strain components---are relaxed to
 minimize the energy.

%% figure showing stress strain curves for three different sized systems.

Fig.~\ref{stressStrainCurves} shows stress-strain curves for shear deformation
simulations for three different system sizes. In all cases there is initially
a smooth, almost linear increase, marking the elastic regime. In the smallest
system, this is first interrupted by small kinks in the stress-strain curve,
and then by sharp drops in stress. The remainder of the curve is characterized
by intervals of almost linear increase punctuated by further sharp drops. The
highest stress attained is prior to the first significant drops. The size of
fluctuations (drops) progressively diminishes for larger system sizes, so that
for the 131072-atom system we observe a quite smooth curve with a well-defined
flow stress following an initial peak.

%% figure showing histograms of sigma_c and slope.

\begin{figure}
\epsfig{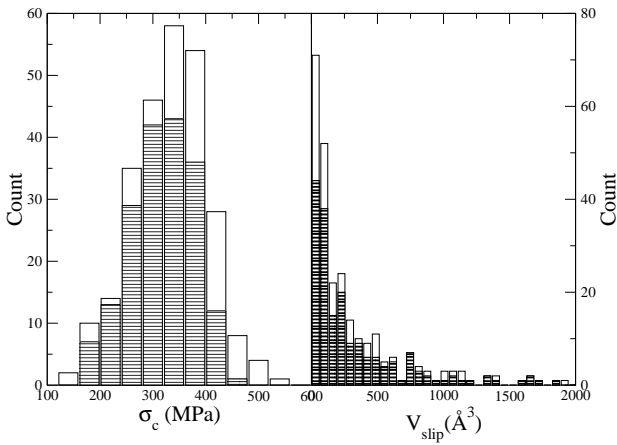}
\caption{\label{distributions}Distribution of $\sigma_c$ and 
$V_{\textrm{slip}}$ values for all events (unshaded) and events taking place
 above 10\% strain (shaded). The means are 331(316) MPa and 
305(317) \AA $^3$, where the figures in parentheses indicate the means of the 
shaded distributions.}
\end{figure}

A natural interpretation of the stress drops in the small system  is 
that they represent discrete plastic flow ``events'', corresponding to
 the postulated STZs of theory or the localized events found in previous 
simulations. The fact that the slope of the stress curve between the drops
is always the same indicates that here only elastic deformation is taking
place and that the plastic deformation is entirely accounted for by processes
associated with the drops. In the following we shall analyze the individual 
events, as identified in the small systems and use this information to
address the properties of the large system. To confirm that the deformation of
the large system also happens through localized events, we have used the 
technique from 
Ref.~\onlinecite{Falk/Langer:1998} to highlight atoms with a large deviation
from affine deformation\footnote{We compare configurations 
differing in strain by 0.1\% and use a cutoff of 10 \AA$^2$, 
consistent with that used in Ref.~\onlinecite{Falk/Langer:1998}.}, and
observing by direct visualization that the atoms so highlighted tend to form 
clusters, have generated statistics of these clusters. A cluster
is a group of (at least three) so-highlighted atoms  
connected by nearest neighbor bonds. We count 5--7 clusters per nm$^3$ per unit
 strain in the ``steady-state''
regime between 10\% and 20\% strain. These clusters typically contain
3--20 atoms, although extreme cases involving up to 125 atoms also occur.
In the small system an event marked by a stress drop typically involves
one or two such clusters, thus  we can use this system to study single events
in detail.

The events are transitions involving internal rearrangements of atoms. We have
identified two characteristic quantities associated with events. The 
first of these, termed ``slip volume'', is geometrical
in nature and represents the amount of plastic strain associated with the
event. The plastic strain is defined in terms of changes in the
shape of the periodic simulation cell. However, if we take the event to be
localized in the cell, the plastic strain induced at the boundaries depends 
not just on the geometry of the rearranging atoms but also on the volume
of the cell. To see this, consider an idealized slip event as a planar
 area $A \hat n$ cut within the material, and the resulting free surfaces 
shifted relatively by an amount $\vec b \perp \hat n$, as in dislocation loop 
nucleation. Then the plastic shear strain felt by the boundary of the system is
$\epsilon_{pl} \sim V_{\text{slip}}/V$, where $V_{\text{slip}} = b A$, and $V$
is the system volume. Alternatively, knowing $\epsilon_{pl}$ we can multiply
by $V$ to obtain $V_{\textrm{slip}}$. This slip volume is 
a tensorial quantity, but in this work we are only concerned with the 
component corresponding to the applied shear strain. We cannot decompose it
 into $\vec b$ and $A \hat n$, since the events are in general geometrically
 more complex than planar slip.

As we describe below we have determined the distribution of $V_{\text{slip}}$,
shown in the right panel of Fig.~\ref{distributions}. Rather than being
peaked at a finite value, the most likely value tends to zero.
The distribution is in fact roughly exponential, although with some extra 
weight at large values: The mean is $\bar V_{\text{slip}} \sim 305$\AA$^3$; if 
an exponential is fitted to the initial data, a slightly smaller
characteristic $V_{\text{slip}}$ of 240 \AA$^3$ is obtained. In a large system,
 the total strain can be written
$\epsilon_{tot} = N_e \bar V_{\text{slip}} / V $, where $N_e$ is the total 
number of events. This implies that 
$\bar V_{\text{slip}}^{-1} = 3.2\text{nm}^{-3}$ is the number density of events
 per unit strain and volume. This is roughly a factor of two smaller than the 
number obtained by counting clusters; this is partly due to the fact that
 in the small system energetically isolated events
may have more than one spatially separate cluster, and partly due to a possible
 enhancement of spatially separate events ``cascading'' together due to 
interactions with periodic images in a small system.
Cascading has been studied by Maloney\cite{Maloney/LeMaitre:2004}. If we
use the estimation of the mean from the exponential fit (which ignores excess
 large events) instead of the actual mean, we find an event density 
$4.2\text{nm}^{-3}$, which is closer to that determined geometrically.

% fitted slope is -0.004230827, with error 0.000483
% corresponds to density 4.2+/- 0.5 events per nm^3 per strain.

% variance of exponential distribution is twice the square of the mean.

The second quantity is a {\em critical stress}, $\sigma_c$, the value of the
shear stress at which the event happens spontaneously at T=0. At
lower stresses, there exists a barrier which might be crossed due to thermal
fluctuations at finite temperature, but at T=0 prevents the event from taking
place until the stress rises high enough. To properly define $\sigma_c$ we 
adopt a stress-controlled formalism, where the six strains all become degrees
of freedom, and a stress term is added to the potential energy, with stress
as a tunable parameter. There then exists a (3N+6)-dimensional potential 
energy (or enthalpy) landscape, which is effectively tilted by increasing the
stress.

For each event apparent in the stress-strain curve, we take configurations from
the simulation before and after the stress drop. These are close to minima of 
the enthalpy landscape. By minimization under a chosen stress we obtain
locally stable configurations which we take as the ``initial'' and ``final''
states for the given event and the given stress. These are used to define
$V_{\textrm{slip}}$. The enthalpy barrier between the events is determined 
using the Nudged Elastic Band 
method\cite{Jonsson/Mills/Jacobsen:1998, Henkelman/Jonsson:2000, 
Henkelman/Uberuaga/Jonsson:2000}. These are computed for a range of 
stresses sufficient to determine $\sigma_c$ with accuracy; they are stress 
dependent, but the dependence is small in the case of $V_{\textrm{slip}}$, 
and was averaged over. We have taken pairs 
of configurations and calculated the above quantities in this way for 
every peak on the stress-strain curve from a shear deformation simulation
of a 2048-atom system up to 30\% strain.

\begin{figure}
\epsfig{file=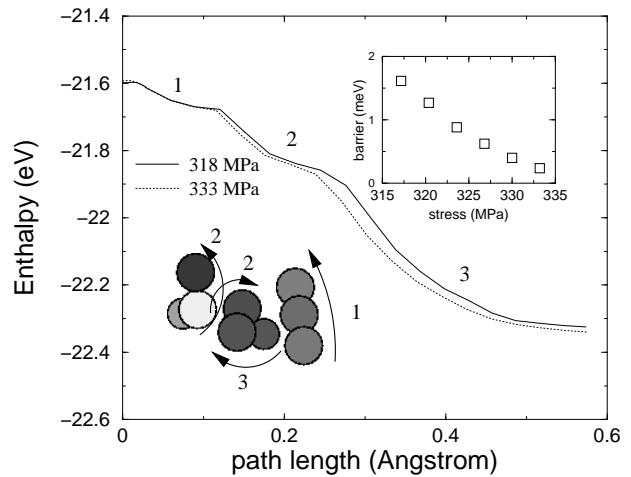,width=3.2 in}
\caption{\label{NEBpaths}Enthalpy along several minimum energy paths for an 
event, for different applied stresses as indicated; the barrier is just visible
as a slight bump at the beginning of the path. In the bottom are shown the
 participating atoms, colored according to deviation from affine deformation
 computed 
between initial and final states, with arrows giving an indication of the types
 of motion and the numbers identifying the order of motions and the 
corresponding features in the enthalpy curve. Inset: stress-dependence of the 
barrier.}
\end{figure}

Enthalpy profiles for a particular event are shown in 
Fig.~\ref{NEBpaths}. In the regime where the calculations have been done, i.e.,
fairly close to the critical stress at which the barrier vanishes (necessary
for events to occur at T=0), the barrier is very small (1--10 meV)
compared to the overall enthalpy change ($\sim$1 eV). Also shown in the 
figure are the atoms
most involved (determined by deviation from affine deformation). An animation 
of the process
shows that the nature and order of the individual motions is as roughly 
indicated by the arrows. This event is relatively simple; some involve
several tens of atoms and complex patterns of motion, although much of this can
 be decomposed into such small pieces involving snake-like motion and 
rotations of groups of three or four atoms.

The inset of Fig.~\ref{NEBpaths} shows the barrier's stress dependence, whose 
form is a steady decrease with increasing stress, flattening
out somewhat at the critical stress $\sigma_c$ where the barrier vanishes.
In fact, the barrier must vanish with a 3/2 power law
sufficiently close to $\sigma_c$, due to the merging of a saddle-point
 and a local minimum of the enthalpy (as in a saddle-node bifurcation). The
 resolution of our data is not enough to identify this; simply using a linear
or quadratic fit to the data near $\sigma_c$ is sufficient to identify
 $\sigma_c$ with an accuracy of $\sim 2$ MPa. The vanishing of the barrier 
(and the minimum) has been studied in detail by Malandro
 and Lacks\cite{Malandro/Lacks:1998, Malandro/Lacks:1999}. At lower stresses, 
there is extra structure in the barrier height, such as abrupt changes of slope
 or even local maxima.

In some cases an intermediate minimum was found along the minimum energy path,
and in these cases separate calculations were made for each of the 
thus-identified ``sub-events''. By combining the results from all barrier 
determinations we can plot the distribution $g(\sigma_c)$ of critical stresses,
 shown in Fig.~\ref{distributions}, left. 
To improve the statistics, we have also included events obtained by 
shear-deformation in the $x-z$ and $y-z$ planes, yielding a total of 262 
events. There is a broad peak, with a mean of 331 MPa and a 
standard deviation of about 70 MPa. If we count only events taking place after
10\% deformation (the shaded distribution in Fig.~\ref{distributions}), the 
mean is a little lower, 316 MPa. This latter value is 
close to the flow stress observed in the large system.

%% figure showing NEB paths for an event
%% THIS IS FROM BAR042

Why is this? To obtain a connection between the deformation behavior of the 
small system and that of the large,
we have computed the stress averaged over a subset of the 131072-atom
 system whose volume is that of the 2048-atom system. This stress-strain curve
is plotted in the lower part of Fig.~\ref{stressStrainCurves}. The striking 
feature of this curve is that it looks closer to that of the actual 
2048-atom system than it does to that
 of the 131072-atom system as a whole, although the stress drops
are not quite as large, nor as sharp. The particular subset was chosen to 
surround a cluster which was active at strain 0.064, and
indeed a significant drop in the stress can be seen at that
strain. The large system thus in a sense behaves as a collection of weakly 
coupled small systems, each
 undergoing relatively large stress fluctuations, but whose average is quite
smooth. The fact that the stress drops are gentler than in the true 2048-atom
system is presumably due the smaller constraints on this region provided
by the surrounding material, compared to those of periodic boundary 
conditions; stress can relax into neighboring material during the
 relaxation to mechanical equilibrium.

Simple considerations provide crude estimates for the mesoscopic flow
stress. In the manner of averaging elastic constants in 
polycrystals\cite{Hirth/Lothe:1982}, we can assume that either the 
stress or the strain is uniform over subsystems. In the first case we imagine
imposing a fixed stress on all subsystems and letting them respond 
independently. Then under relaxation, every subsystem will flow until it 
reaches the first critical stress that is higher than the imposed stress. No 
further deformation can take place---unless the imposed
stress is higher than the maximum critical stress. Thus the flow stress
 is the maximum of the $\sigma_c$-distribution. This is at least an upper 
bound: The material clearly cannot sustain a stress greater than
 the maximum $\sigma_c$, 450 MPa for events taking place above 10\% strain.

Alternatively, imposing a uniform strain on each subsystem, each undergoes 
deformation just like the single simulated small system. Assuming the 
individual stress-strain curves have no fixed phase relation, averaging
 across them at a given strain is equivalent to averaging over the strain 
 history of a single subsystem. This average is straightforward to compute if
the assumption of a fixed $V_{\textrm{slip}}$ is made, yielding 
$\bar \sigma_c - \Delta \sigma /2$, where $\Delta 
\sigma=2\mu V_{\textrm{slip}}/V_{sub}$ is the stress drop
associated with individual events in a subsystem. This value is 265 MPa with
our choice of small system ($V_{sub}=4.3\times10^4$\AA$^3$), about 10\%
less than the observed flow stress.

This estimate is necessarily rough, in particular because it explicitly
involves an apparently arbitrary subsystem size. However, the size of our
small system is not so arbitrary. We have previously noted that it corresponds
more or less to the size at which events become discrete. Furthermore, analysis
of the distribution of stress (not shown here) averaged over various-sized
 subsystems suggests that this size is about the smallest at which random
fluctuations coming from the atomic stresses begin to cancel out enough to make
 the averaged stress meaningful---a quantity which actually represents force
per unit area exerted by the material on itself. Apart from these 
considerations, the directly observed correspondence between the flow stress
and the mean $\sigma_c$ itself supports the overall picture of the statistics
of small subsystems determining the mesoscopic plastic behavior.

\begin{acknowledgments}
This work was supported by the Danish Research Councils through
Grant No. 5020-00-0012 and by the Danish Center for Scientific
Computing through Grant No. HDW-1101-05. 
\end{acknowledgments}

% no spaces between database names!
%\bibliography{glass,glassMechanics,extra}

\begin{thebibliography}{20}
\expandafter\ifx\csname natexlab\endcsname\relax\def\natexlab#1{#1}\fi
\expandafter\ifx\csname bibnamefont\endcsname\relax
  \def\bibnamefont#1{#1}\fi
\expandafter\ifx\csname bibfnamefont\endcsname\relax
  \def\bibfnamefont#1{#1}\fi
\expandafter\ifx\csname citenamefont\endcsname\relax
  \def\citenamefont#1{#1}\fi
\expandafter\ifx\csname url\endcsname\relax
  \def\url#1{\texttt{#1}}\fi
\expandafter\ifx\csname urlprefix\endcsname\relax\def\urlprefix{URL }\fi
\providecommand{\bibinfo}[2]{#2}
\providecommand{\eprint}[2][]{\url{#2}}

\bibitem[{\citenamefont{Greer}(1995)}]{Greer:1995}
\bibinfo{author}{\bibfnamefont{A.~L.} \bibnamefont{Greer}},
  \bibinfo{journal}{Science} \textbf{\bibinfo{volume}{267}},
  \bibinfo{pages}{1947} (\bibinfo{year}{1995}).

\bibitem[{\citenamefont{Johnson}(1999)}]{Johnson:1999}
\bibinfo{author}{\bibfnamefont{W.~L.} \bibnamefont{Johnson}},
  \bibinfo{journal}{MRS Bulletin} \textbf{\bibinfo{volume}{24}},
  \bibinfo{pages}{42} (\bibinfo{year}{1999}).

\bibitem[{\citenamefont{Khonik et~al.}(1998)\citenamefont{Khonik, Kosilov,
  Mikhailov, and Sviridov}}]{Khonik/others:1998}
\bibinfo{author}{\bibfnamefont{V.~A.} \bibnamefont{Khonik}},
  \bibinfo{author}{\bibfnamefont{A.~T.} \bibnamefont{Kosilov}},
  \bibinfo{author}{\bibfnamefont{V.~A.} \bibnamefont{Mikhailov}},
  \bibnamefont{and} \bibinfo{author}{\bibfnamefont{V.~V.}
  \bibnamefont{Sviridov}}, \bibinfo{journal}{Acta. Mater.}
  \textbf{\bibinfo{volume}{46}}, \bibinfo{pages}{3399} (\bibinfo{year}{1998}).

\bibitem[{\citenamefont{Langer et~al.}(2003)\citenamefont{Langer, Pechenik, and
  Falk}}]{Langer/Pechenik/Falk:2003}
\bibinfo{author}{\bibfnamefont{J.~S.} \bibnamefont{Langer}},
  \bibinfo{author}{\bibfnamefont{L.}~\bibnamefont{Pechenik}}, \bibnamefont{and}
  \bibinfo{author}{\bibfnamefont{M.~L.} \bibnamefont{Falk}}
  (\bibinfo{year}{2003}), \eprint{cond-mat/0311057}.

\bibitem[{\citenamefont{Falk and Langer}(1998)}]{Falk/Langer:1998}
\bibinfo{author}{\bibfnamefont{M.~L.} \bibnamefont{Falk}} \bibnamefont{and}
  \bibinfo{author}{\bibfnamefont{J.~S.} \bibnamefont{Langer}},
  \bibinfo{journal}{Phys. Rev. E} \textbf{\bibinfo{volume}{57}},
  \bibinfo{pages}{7192} (\bibinfo{year}{1998}).

\bibitem[{\citenamefont{Srolovitz et~al.}(1982)\citenamefont{Srolovitz, Vitek,
  and Egami}}]{Srolovitz/Vitek/Egami:1982}
\bibinfo{author}{\bibfnamefont{D.}~\bibnamefont{Srolovitz}},
  \bibinfo{author}{\bibfnamefont{V.}~\bibnamefont{Vitek}}, \bibnamefont{and}
  \bibinfo{author}{\bibfnamefont{T.}~\bibnamefont{Egami}},
  \bibinfo{journal}{Acta. Metall.} \textbf{\bibinfo{volume}{31}},
  \bibinfo{pages}{335} (\bibinfo{year}{1982}).

\bibitem[{\citenamefont{Maeda and Takeuchi}(1981)}]{Maeda/Takeuchi:1981}
\bibinfo{author}{\bibfnamefont{K.}~\bibnamefont{Maeda}} \bibnamefont{and}
  \bibinfo{author}{\bibfnamefont{S.}~\bibnamefont{Takeuchi}},
  \bibinfo{journal}{Philos. Mag. A} \textbf{\bibinfo{volume}{44}},
  \bibinfo{pages}{643} (\bibinfo{year}{1981}).

\bibitem[{\citenamefont{Malandro and Lacks}(1999)}]{Malandro/Lacks:1999}
\bibinfo{author}{\bibfnamefont{D.~L.} \bibnamefont{Malandro}} \bibnamefont{and}
  \bibinfo{author}{\bibfnamefont{D.~J.} \bibnamefont{Lacks}},
  \bibinfo{journal}{J. Chem. Phys.} \textbf{\bibinfo{volume}{110}},
  \bibinfo{pages}{4593} (\bibinfo{year}{1999}).

\bibitem[{\citenamefont{Lund and Schuh}(2003{\natexlab{a}})}]{Lund/Schuh:2003b}
\bibinfo{author}{\bibfnamefont{A.~C.} \bibnamefont{Lund}} \bibnamefont{and}
  \bibinfo{author}{\bibfnamefont{C.~A.} \bibnamefont{Schuh}},
  \bibinfo{journal}{Acta Mater.} \textbf{\bibinfo{volume}{51}},
  \bibinfo{pages}{5399} (\bibinfo{year}{2003}{\natexlab{a}}).

\bibitem[{\citenamefont{Malandro and Lacks}(1998)}]{Malandro/Lacks:1998}
\bibinfo{author}{\bibfnamefont{D.~L.} \bibnamefont{Malandro}} \bibnamefont{and}
  \bibinfo{author}{\bibfnamefont{D.~J.} \bibnamefont{Lacks}},
  \bibinfo{journal}{Phys. Rev. Lett.} \textbf{\bibinfo{volume}{81}},
  \bibinfo{pages}{5576} (\bibinfo{year}{1998}).

\bibitem[{\citenamefont{Lacks}(2001)}]{Lacks:2001}
\bibinfo{author}{\bibfnamefont{D.~J.} \bibnamefont{Lacks}},
  \bibinfo{journal}{Phys. Rev. Lett.} \textbf{\bibinfo{volume}{87}},
  \bibinfo{pages}{225502} (\bibinfo{year}{2001}).

\bibitem[{\citenamefont{Lund and Schuh}(2003{\natexlab{b}})}]{Lund/Schuh:2003a}
\bibinfo{author}{\bibfnamefont{A.~C.} \bibnamefont{Lund}} \bibnamefont{and}
  \bibinfo{author}{\bibfnamefont{C.~A.} \bibnamefont{Schuh}},
  \bibinfo{journal}{Nature Mater.} \textbf{\bibinfo{volume}{2}},
  \bibinfo{pages}{449} (\bibinfo{year}{2003}{\natexlab{b}}).

\bibitem[{\citenamefont{Sommer et~al.}(1980)\citenamefont{Sommer, Bucher, and
  Predal}}]{Sommer/Bucher/Predal:1980}
\bibinfo{author}{\bibfnamefont{F.}~\bibnamefont{Sommer}},
  \bibinfo{author}{\bibfnamefont{G.}~\bibnamefont{Bucher}}, \bibnamefont{and}
  \bibinfo{author}{\bibfnamefont{B.}~\bibnamefont{Predal}},
  \bibinfo{journal}{J. Phys. Colloque C8} \textbf{\bibinfo{volume}{41}},
  \bibinfo{pages}{563} (\bibinfo{year}{1980}).

\bibitem[{\citenamefont{Jacobsen et~al.}(1996)\citenamefont{Jacobsen, Stoltze,
  and N{\o}rskov}}]{Jacobsen/Stoltze/Norskov:1996}
\bibinfo{author}{\bibfnamefont{K.~W.} \bibnamefont{Jacobsen}},
  \bibinfo{author}{\bibfnamefont{P.}~\bibnamefont{Stoltze}}, \bibnamefont{and}
  \bibinfo{author}{\bibfnamefont{J.~K.} \bibnamefont{N{\o}rskov}},
  \bibinfo{journal}{Surf. Sci.} \textbf{\bibinfo{volume}{366}},
  \bibinfo{pages}{394} (\bibinfo{year}{1996}).

\bibitem[{\citenamefont{Bailey et~al.}(2004)\citenamefont{Bailey, Schi{\o}tz,
  and Jacobsen}}]{Bailey/Schiotz/Jacobsen:2004}
\bibinfo{author}{\bibfnamefont{N.~P.} \bibnamefont{Bailey}},
  \bibinfo{author}{\bibfnamefont{J.}~\bibnamefont{Schi{\o}tz}},
  \bibnamefont{and} \bibinfo{author}{\bibfnamefont{K.~W.}
  \bibnamefont{Jacobsen}}, \bibinfo{journal}{Phys. Rev. B}
  \textbf{\bibinfo{volume}{69}}, \bibinfo{pages}{144205}
  (\bibinfo{year}{2004}).

\bibitem[{\citenamefont{Maloney and
  LeMa\^{i}tre}(2004)}]{Maloney/LeMaitre:2004}
\bibinfo{author}{\bibfnamefont{C.}~\bibnamefont{Maloney}} \bibnamefont{and}
  \bibinfo{author}{\bibfnamefont{A.}~\bibnamefont{LeMa\^{i}tre}}
  (\bibinfo{year}{2004}), \eprint{cond-mat/0402148}.

\bibitem[{\citenamefont{J\'{o}nsson et~al.}(1998)\citenamefont{J\'{o}nsson,
  Mills, and Jacobsen}}]{Jonsson/Mills/Jacobsen:1998}
\bibinfo{author}{\bibfnamefont{H.}~\bibnamefont{J\'{o}nsson}},
  \bibinfo{author}{\bibfnamefont{G.}~\bibnamefont{Mills}}, \bibnamefont{and}
  \bibinfo{author}{\bibfnamefont{K.~W.} \bibnamefont{Jacobsen}}, in
  \emph{\bibinfo{booktitle}{Classical and Quantum Dynamics in Condensed Phase
  Simulations}}, edited by \bibinfo{editor}{\bibfnamefont{B.~J.}
  \bibnamefont{Berne}},
  \bibinfo{editor}{\bibfnamefont{G.}~\bibnamefont{Ciccotti}}, \bibnamefont{and}
  \bibinfo{editor}{\bibfnamefont{D.~F.} \bibnamefont{Coker}}
  (\bibinfo{publisher}{World Scientific}, \bibinfo{year}{1998}).

\bibitem[{\citenamefont{Henkelman and
  J\'{o}nsson}(2000)}]{Henkelman/Jonsson:2000}
\bibinfo{author}{\bibfnamefont{G.}~\bibnamefont{Henkelman}} \bibnamefont{and}
  \bibinfo{author}{\bibfnamefont{H.}~\bibnamefont{J\'{o}nsson}},
  \bibinfo{journal}{J. Chem. Phys.} \textbf{\bibinfo{volume}{113}},
  \bibinfo{pages}{9978} (\bibinfo{year}{2000}).

\bibitem[{\citenamefont{Henkelman et~al.}(2000)\citenamefont{Henkelman,
  Uberuaga, and J\'{o}nsson}}]{Henkelman/Uberuaga/Jonsson:2000}
\bibinfo{author}{\bibfnamefont{G.}~\bibnamefont{Henkelman}},
  \bibinfo{author}{\bibfnamefont{B.}~\bibnamefont{Uberuaga}}, \bibnamefont{and}
  \bibinfo{author}{\bibfnamefont{H.}~\bibnamefont{J\'{o}nsson}},
  \bibinfo{journal}{J. Chem. Phys.} \textbf{\bibinfo{volume}{113}},
  \bibinfo{pages}{9901} (\bibinfo{year}{2000}).

\bibitem[{\citenamefont{Hirth and Lothe}(1982)}]{Hirth/Lothe:1982}
\bibinfo{author}{\bibfnamefont{J.~P.} \bibnamefont{Hirth}} \bibnamefont{and}
  \bibinfo{author}{\bibfnamefont{J.}~\bibnamefont{Lothe}},
  \emph{\bibinfo{title}{Theory of Dislocations}} (\bibinfo{publisher}{Krieger
  Publishing Company}, \bibinfo{year}{1982}), \bibinfo{edition}{2nd} ed.

\end{thebibliography}

\end{document}